\documentstyle[prl,twocolumn,aps,psfig]{revtex}

\draft
\begin{document}
\twocolumn[\hsize\textwidth\columnwidth\hsize\csname@twocolumnfalse%
\endcsname

\title{On possible superconductivity in the doped ladder compound  
La$_{1-x}$Sr$_x$CuO$_{2.5}$}

\author{B. Normand} 

\address{Departement f\"ur Physik und Astronomie, Universit\"at Basel, 
CH-4056 Basel, Switzerland. }

\author{D. F. Agterberg\cite{pa} and T. M. Rice}

\address{Theoretische Physik, ETH-H\"onggerberg, CH-8093 Z\"urich, 
Switzerland.}

\date{\today}

\maketitle

\begin{abstract}

LaCuO$_{2.5}$ is a system of coupled, two-chain, cuprate ladders 
which may be doped systematically by Sr substitution. Motivated by the 
recent synthesis of single crystals, we investigate theoretically the 
possibility of superconductivity in this compound. We use 
a model of spin fluctuation-mediated superconductivity, where 
the pairing potential is strongly peaked at $\pi$ in the ladder 
direction. We solve the coupled gap equations on the bonding and 
antibonding ladder bands to find superconducting solutions across the 
range of doping, and discuss their relevance to the real material.

\end{abstract}

]


	In the search for the microscopic origin of high-$T_c$ 
superconductivity (HTS), it is essential to understand not only the materials 
in which it occurs, but also why it fails to occur in certain systems which 
are ostensibly only marginally different. In this letter we examine the 
cuprate La$_{1-x}$Sr$_x$CuO$_{2.5}$, a coupled-chain analog of planar 
La$_{2-x}$Sr$_x$CuO$_{4}$. The parent compound was at first thought to 
be a dopable, two-chain spin ladder system \cite{rht}, but it was later shown 
\cite{rmkiahkt,rnr1,rtzu,rnr2} that interladder coupling in this unfrustrated 
structure was sufficiently strong to stabilize long-range antiferromagnetic 
(AF) order. 

	The first doping experiments \cite{rht,rh} were on ceramic samples, 
and found metallic behavior only for relatively high doping values $x \ge 
0.2$. No superconductivity was found for any doping, a result attributed to 
quasi-one-dimensionality (1d) and the unavoidably strong random potential
\cite{rnr1}. Recently, Takagi and coworkers \cite{rmnt} have successfully 
grown single crystals with the fixed composition $x$ = 0.15. These show good 
metallic characteristics, both along and across the ladder direction with 
anisotropy ratio $\rho_{\perp}/\rho_{\parallel} \sim 8$, and residual 
resistivities as low as 40$\mu\Omega {\rm cm}^{-1}$. Although the latter 
value corresponds to a mean free path exceeding 10 ladder unit cells, at 
least comparable to coherence lengths in cuprates, there was no indication 
of superconductivity. The subtlety of the HTS phenomenon was further 
highlighted by the discovery \cite{runatmk} of superconductivity under 
pressure in a different cuprate ladder system. 

	The resistivity data \cite{rmnt} appear to support an anisotropic, 
3d approach. Here we consider in detail the Fermi surface of La$_{1-
x}$Sr$_x$CuO$_{2.5}$, to find out if there are inherent frustrations present 
which could hinder or prevent a superconducting state with the predominantly 
$d$-symmetric character expected in ladders \cite{rdr}. To this end we use 
the model of Millis, Monien and Pines (MMP) \cite{rmmp} for superconductivity 
mediated by AF spin fluctuations, and perform exploratory calculations 
examining the form of the resulting gap function. 


	Considering the structure as two-chain ladders along ${\hat z}$, 
with weaker interladder couplings in the transverse directions, we compute 
tight-binding bands analogous to the bonding and antibonding combinations 
in an isolated ladder. In Ref. \cite{rnr1}, a set of near-neighbor 
hopping parameters was deduced from a four-band fit to the band 
structure computed in the local density approximation (LDA) \cite{rm}. 
Using local, Cu-based orbitals $\phi_1$ and $\phi_2$ on each atom in the 
reduced unit cell, the Hamiltonian in terms of reciprocal-space 
wave functions $\phi_{\mu} ({\bf k}) = V^{-1/2} \sum_i \phi_{\mu} ({\bf r}_i) 
e^{i {\bf k \cdot r}_i}$ ($\mu = 1,2$) is 
\begin{equation} 
H_t = ( \phi_{1}^*, \phi_{2}^* ) \left( \begin{array}{cc} 
t_{11} & t_{12} \\ t_{21} & t_{22} \end{array} \right) \left( 
\begin{array}{c} \phi_1 \\ \phi_2 \end{array} \right) ,
\label{elh}
\end{equation}
where 
\begin{eqnarray}
t_{11} & = & t_{22} = - 2 t_z \cos k_z - 2 t_{z}^{\prime} \cos 2 k_z , 
\label{elhc1} \\ t_{12} & = & t_{21}^* = - {\tilde t}_a - {\tilde t}_b 
[ e^{i k_x} + e^{i k_y} ], \label{elhc2}
\end{eqnarray}
${\tilde t}_i$ denotes $t_i + 2 t_{i}^{\prime} \cos k_z$, the ladder 
lattice constant $a_c$ has been set to 1, and the parameters $t_i$, 
shown in Fig. 1, have the values \cite{rnr1} $t_z = t_a = t = 0.4$eV, 
$t_{z}^{\prime} = \, $\mbox{$\frac{1}{6}$}$ t_z$, $t_{a}^{\prime} = 
- $\mbox{$\frac{1}{5}$}$ t_a$, $t_b = \, $\mbox{$\frac{2}{5}$}$t_a$ and 
$t_{b}^{\prime} = \, $\mbox{$\frac{1}{5}$}$ t_b$. 

\begin{figure}[hp]
\centerline{\psfig{figure=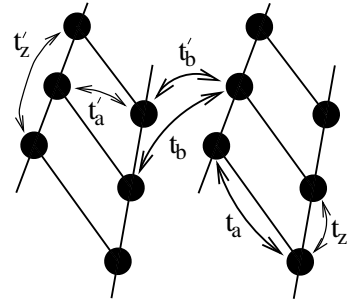,height=4cm,angle=0}}
\medskip
\caption{Two-chain ladder structure of Cu ions in La$_{1-x}$Sr$_x$CuO$_{2.5}$,
showing tight-binding parameters.}
\end{figure}

	The two eigenmodes are $\epsilon_{{\bf k} 
\pm} = t_{11} ({\bf k}) \pm | t_{12} ({\bf k}) |$. 
The nature of the corresponding eigenfunctions is seen 
most clearly by introducing the bonding and antibonding basis functions 
$\phi_{b,a} ({\bf k}) = [ \phi_1 ({\bf k}) \pm \phi_2 ({\bf k}) ] / 
\sqrt{2}$, in terms of which the eigenfunctions are 
\begin{equation}
\phi^{\pm} ({\bf k}) = \cos \theta_{\bf k} \phi_{a,b} ({\bf k}) + i \sin 
\theta_{\bf k} \phi_{b,a} ({\bf k}) \label{eef} ,
\end{equation}
with the phase factor $\theta_{\bf k} = $\mbox{$\frac{1}{2}$}$ {\rm arg} 
(t_{12})$. At the zone center these wave functions have full bonding or 
antibonding character, which is exchanged at the $(\pi,\pi,0)$ point. 
Elsewhere in the zone the character is mixed, except along the line 
$k_x = - k_y$, where the argument remains zero and a band crossing occurs. 
The nature of the eigenfunctions changes little with $k_z$, except close to 
$(\pi,\pi,0)$ where there is another crossing. The two-band dispersion for 
a doping of $x$ = 0.15 is illustrated in Fig. 2(a) for a ${\bf k}$-space 
path equivalent to that depicted in Refs. \cite{rm} and \cite{rnr1}. Although 
no crossings occur on this path, the character of the associated 
eigenfunctions is altered by evolution of the complex phase factor. 
The Fermi surfaces (Fig. 2(b)) for the two bands remain quasi-1d \cite{rnr1}, 
in that they take the form of sheets crossed only along $k_z$, with some 
variation of $k_{z {\rm F}}$ depending on ${\tilde {\bf k}} \equiv (k_x,k_y)$. 

\begin{figure}[hp]
\centerline{\psfig{figure=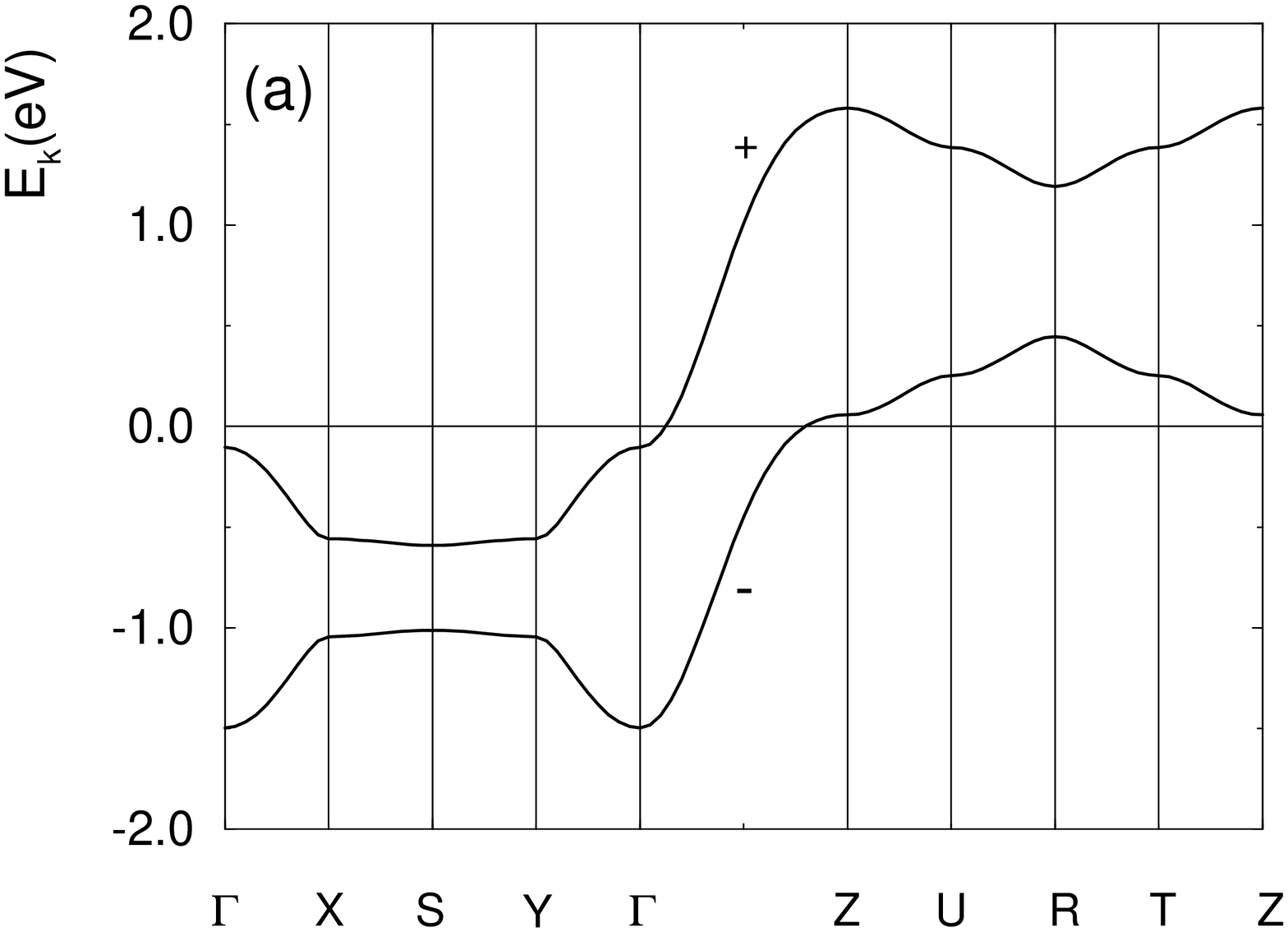,height=5.5cm,angle=270}}
\centerline{\psfig{figure=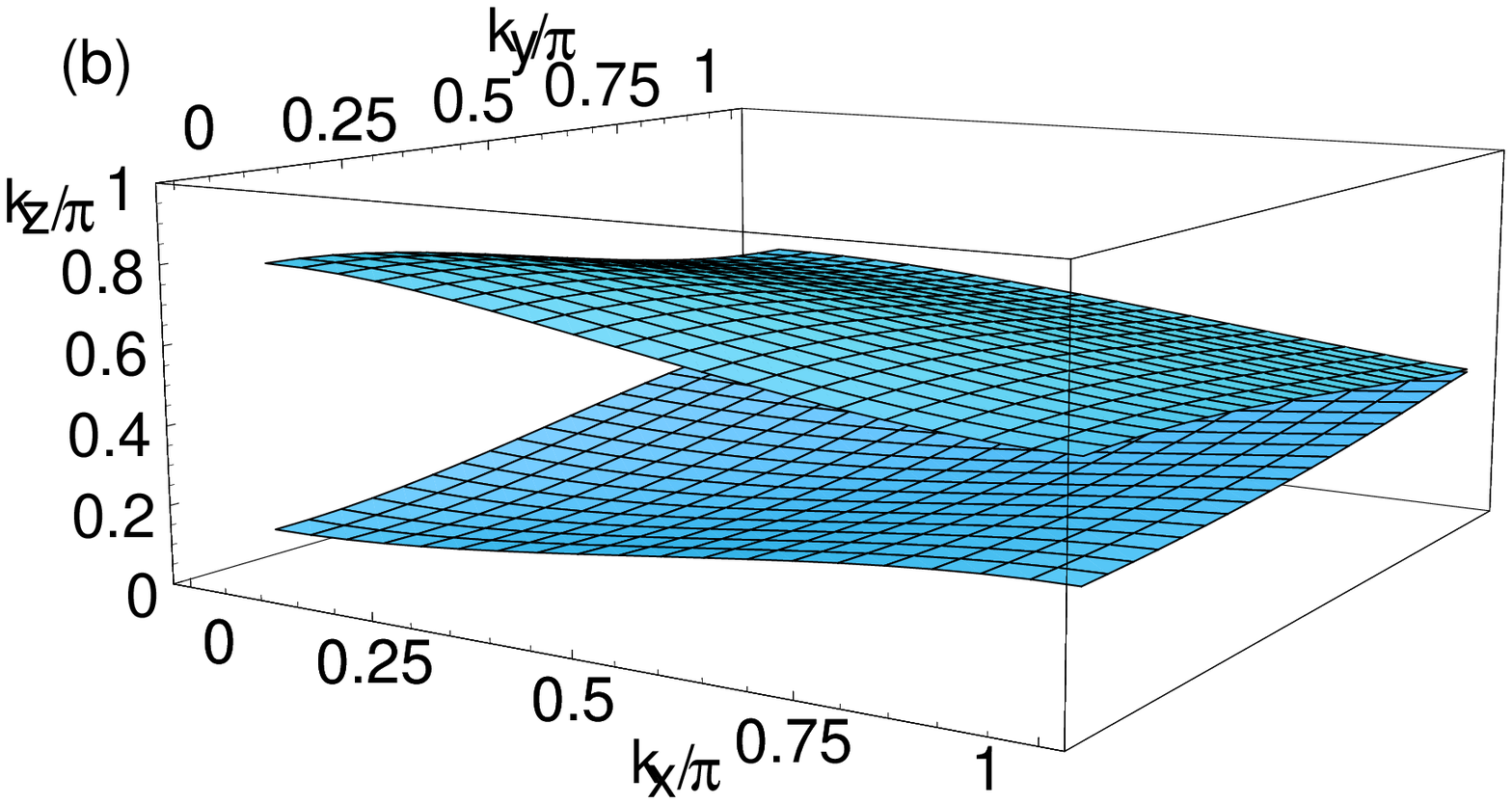,height=4.5cm,angle=0}}
\medskip
\caption{(a) Tight-binding dispersion relations and (b) Fermi surfaces 
for two-band model at doping $x$ = 0.15. }
\end{figure}


	The gap equations describing superconductivity in a two-band 
system take the form  
\begin{equation} 
\Delta_{{\bf k} \alpha} = \frac{1}{V} \sum_{{\bf k}^{\prime},\beta} V_{{\bf k}
{\bf k}^{\prime} \alpha \beta} \frac{\Delta_{{\bf k}^{\prime} \beta}}
{2 E_{{\bf k}^{\prime} \beta}} \tanh \left( \frac {E_{{\bf k}^{\prime} 
\beta}}{2 k_B T} \right) ,
\label{egge}
\end{equation}
where $\alpha,\beta = \pm$ are the band indices, $V_{{\bf k} {\bf k}^{\prime} 
\alpha \beta}$ is a matrix containing both the on-site repulsion term and 
the pairing potential, and $E_{{\bf k} \alpha} = \pm \sqrt{(\epsilon_{{\bf k} 
\alpha} - \mu)^2 + \Delta_{{\bf k} \alpha}^2}$ specifies the dispersion 
relations in the superconducting state ($\mu$ is the chemical potential). 
The symmetry of the gaps $\Delta_{{\bf k} \pm}$, both along ($k_z$) and 
transverse (${\tilde {\bf k}}$) to the ladder axis, is of particular interest.

	The pairing potential, as usual in cuprates, is repulsive. Scattering 
processes due to interactions with spin fluctuations should be strongly 
enhanced at the characteristic wave vector of the AF correlated ladder 
spins, $q_z = \pi$. The dominant contributions are expected from interband 
scattering of intraband electron pairs when the band separation is close 
to $\pi$, as represented schematically in Fig 3. To describe this 
we adopt the frequency-dependent MMP potential in the form 
\begin{equation} 
V_{{\bf k} {\bf k}^{\prime} \alpha \beta} = \frac{ - V_0 \epsilon_{\alpha 
\beta}}{\sqrt{\left( 1 + \xi_z^2 |q_z - Q_z|^2 \right)^2 + \left( \omega^2 
/ \omega_{\rm SF}^2 \right)}} .
\label{evmmp}
\end{equation}
Here $q_z = k_z - k_{z}^{\prime}$, $Q_z = \pi$ is the AF wave 
vector in the ladder direction, $\omega = \epsilon_{\bf k} - 
\epsilon_{{\bf k}^{\prime}}$ and $\omega_{\rm SF}$ sets the frequency 
scale. $\epsilon_{\alpha \beta}$ is the off-diagonal tensor ensuring 
the interband scattering condition, and has the effect of a projector, 
$\sum_{m \ne n} |\phi_m \rangle \langle \phi_n|$, $m,n = a,b$, restricting 
the processes of interest to those scattering pairs between bonding and 
antibonding states. The longitudinal correlation length $\xi_z$ determines 
the spin fluctuation enhancement, and may be deduced from the MMP parameters 
\cite{rmmp,rmp} for planar cuprate superconductors. We neglect the transverse 
dispersion of spin fluctuations: because \cite{rtzu,rnr2} interladder 
superexchange is weak ($J_{\perp} \simeq 0.12 J$), this would contribute 
an effect of order 10\%, with the consequence that the leading components 
of any gap solution will be isotropic in ${\tilde {\bf k}}$. 

\begin{figure}[hp]
\centerline{\psfig{figure=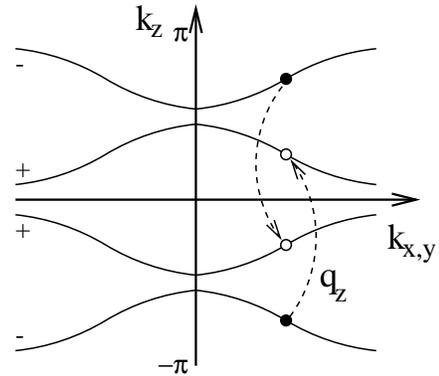,height=5.0cm,angle=0}}
\medskip
\caption{Schematic Fermi-surface geometry showing important interband 
pair scattering processes. }
\end{figure}

	The Hubbard-type model under consideration has also a repulsive 
on-site interaction $U$, which appears in the Hamiltonian as $H_{\rm U} = 
$\mbox{$\frac{1}{2}$}$ U \sum_{{\bf k} \alpha = \pm} ( \Delta_{{\bf k} 
\alpha} c_{\bf k \uparrow}^{\alpha \dag} c_{-{\bf k} \downarrow}^{\alpha 
\dag} + H. c.)$. In the 
appropriate ($t$-$J$) limit of $U \rightarrow \infty$, the amplitude for 
pair creation on any site must be zero, which sets the constraint 
$\sum_{\bf k} c_{\bf k \uparrow}^{+ \dag} c_{-{\bf k} \downarrow}^{+ \dag} + 
c_{\bf k \uparrow}^{- \dag} c_{-{\bf k} \downarrow}^{- \dag} = 0$, or 
\begin{equation}
\sum_{\bf k} \left[ \frac{\Delta_{{\bf k} +}}{2 E_{{\bf k} +}} + 
\frac{\Delta_{{\bf k} -}}{2 E_{{\bf k} -}} \right] = 0. 
\label{egce}
\end{equation}
The full, coupled gap equations can be written as 
\begin{equation}
\Delta_{{\bf k} \alpha} = - \frac{1}{V} \sum_{{\bf k}^\prime} \left[ 
\frac{U}{2} D_{{\bf k}^{\prime} \alpha} + \left( \frac{U}{2} + 
\frac{V_0}{B} \right) D_{{\bf k}^{\prime} \bar{\alpha}} \right] , 
\label{encge} 
\end{equation}
where 
\begin{equation}
D_{{\bf k} \alpha} = \frac{\Delta_{{\bf k} \alpha}} {2 E_{{\bf k} 
\alpha}} \tanh \left( \frac {E_{{\bf k} \alpha}} {2 k_B T} \right) , 
\label{echi}
\end{equation}
$\alpha = \pm$, $\bar{\alpha} \equiv - \alpha$, and $B$ denotes the 
denominator in Eq. (\ref{evmmp}). As both $U$ and $V$ terms are 
repulsive, the most straightforward solution is one where the gaps 
have opposite signs on the $\pm$ bands. The symmetry of the $U$ terms 
in $D_+$ and $D_-$ ensures that the constraint (\ref{egce}) is obeyed.

	We take the parameter values from planar cuprates\cite{rmp}, 
where the local physics can be expected to be similar.
The correlation length is $\xi_z = 2.3a_c$. The prefactor is $V_0 = 
g^2 \chi_{\rm Q}$, where $g$ = 1.53eV is the coefficient of the electron-spin 
fluctuation interaction and $\chi_{\rm Q} = \pi (\xi/a)^2 \chi_0$ is a wave 
vector-adjusted static susceptibility; $\chi_0$ = 2.62 states/eV gives 
$V_0 \simeq 100$. The damping frequency is given by $\omega_{\rm SF} = 
\Gamma / (\pi \xi / a)^2$, where $\Gamma$ = 0.4eV is the frequency cutoff  
of the interaction, whence $\omega_{\rm SF}$ = 8meV. The values of all 
parameters are taken to be constant at and below $T_c$. Considering the 
range of $U$, the constraint (\ref{egce}) is satisfied to within 
1\% for $U = 10J$, so this value is used below. 

	Solution of the gap equations (\ref{encge}) is simplified 
by separating the dependences of the pairing potential on $k_z$ and 
${\tilde {\bf k}}$, and a harmonic decomposition of the gap functions 
\begin{equation}
\Delta_{\bf k}^{\pm} = \Delta_{0}^{\pm} + \Delta_{1}^{\pm} \cos k_z + 
\Delta_{2}^{\pm} \cos 2 k_z + \dots 
\label{eghd}
\end{equation} 
and the potential 
\begin{equation}
V({\bf k},k_{z}^{\prime}) = \sum_{ij} V_{ij} ({\tilde {\bf k}}) 
\cos i k_z \cos j k_{z}^{\prime} .
\label{epphd}
\end{equation}
The dominant contributions to $V_{ij} ({\tilde {\bf k}})$ are from 
Fermi-surface scattering processes with $\epsilon_{\bf k} = 0$. The 
approximate treatment is justified by the parameters $\xi_z$ and 
$\omega_{\rm SF}$, which are such that the halfwidth of $V(q_z)$ is 
$\delta q \sim 0.2\pi$, a quantity exceeding the variation in band separation 
$k_{z {\rm F}}^+ - k_{z {\rm F}}^-$ across the Brillouin zone (Fig. 
2(b)). Thus scattering vectors of all ${\tilde {\bf q}}$ contribute 
similarly, and the averaging performed in the approximation introduces 
insignificant errors.

	The qualitative nature of the solutions is influenced by 
the doping. In this treatment the band shape is not affected, but 
a change of doping alters the separation of the Fermi surfaces 
(Fig. 3). The pairing is (broadly) peaked where the bands have maximal 
areas separated by $q_z \simeq \pi$. We find that at $x_m = 0.112$ 
the separation is $\pi$ at ${\tilde {\bf k}}$ = (0,0), and is smaller 
than $\pi$ everywhere else in the zone. Increased doping ($x > x_m$) 
causes the net interaction to diminish, while at $x < x_m$ a circle of 
band separation $\pi$ will open and increase in circumference, implying a 
monotonic rise in any $T_c$, at least until $x = 0.082$ where Fermi 
surface is lost to formation of an electron pocket at $Z$ in the $-$ band. 
We have solved the gap equations for a variety of doping values $0.05 < 
x < 0.2$, retaining interaction coefficients $V_{ij}$ up to fourth order, 
and gap components $\Delta_{i}$ to $i$ = 2. 

\begin{figure}[hp]
\centerline{\psfig{figure=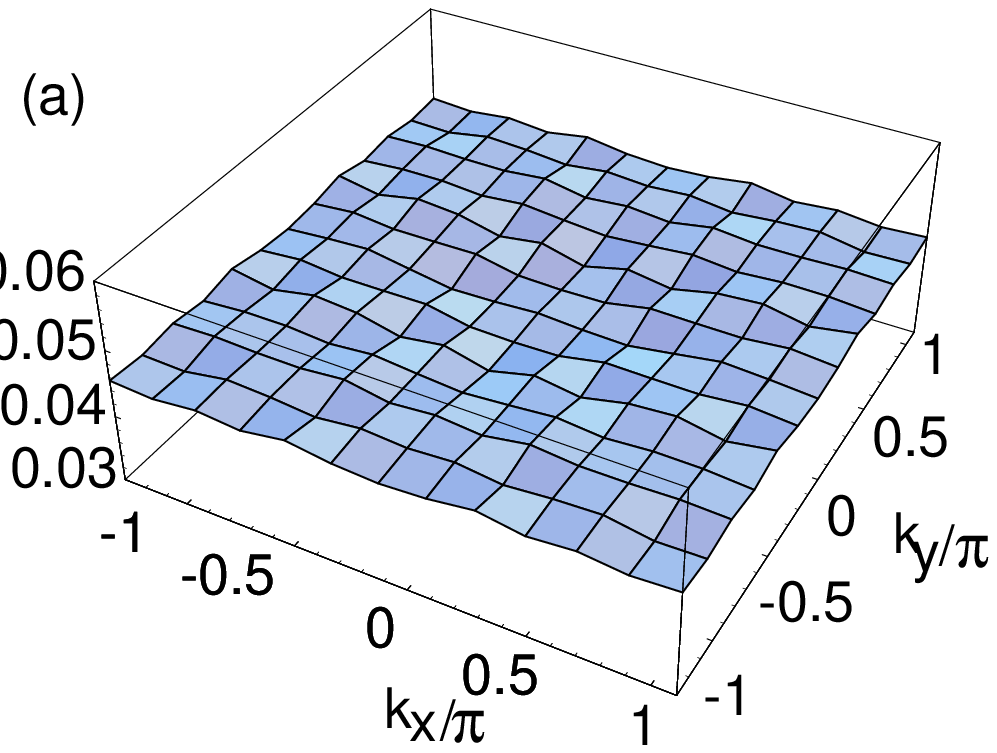,height=5.0cm,angle=0}}
\centerline{\psfig{figure=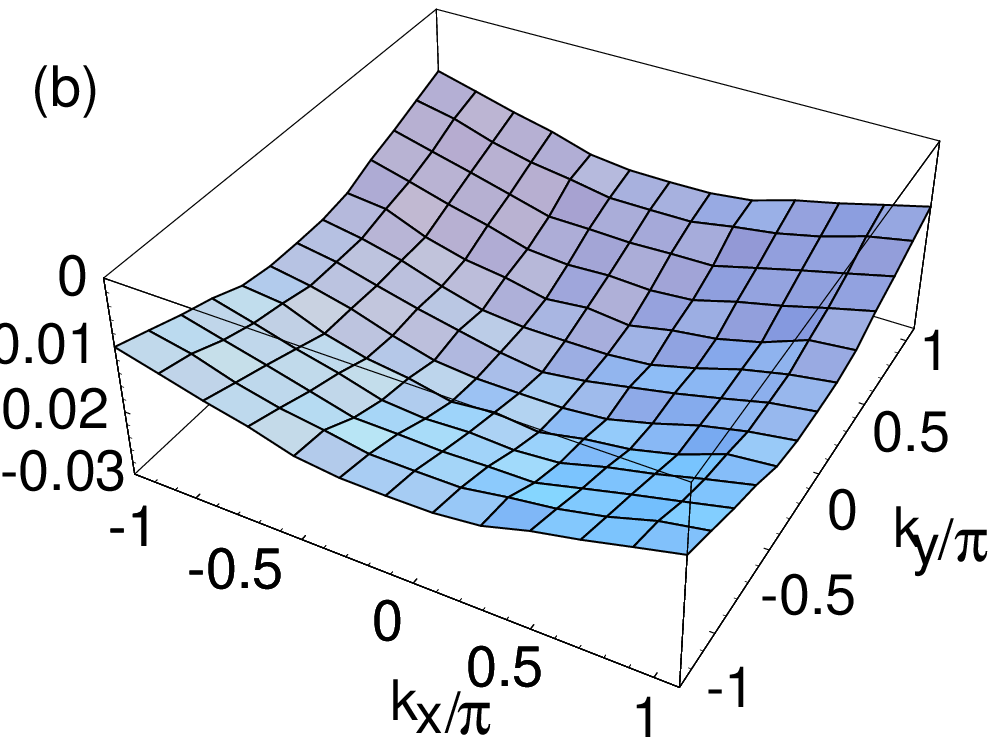,height=5.0cm,angle=0}}
\medskip
\caption{Superconducting gaps (a) $\Delta_{\tilde {\bf k}}^+$ and (b) 
$\Delta_{\tilde {\bf k}}^-$ on the Fermi surfaces for doping $x$ = 0.082. }
\end{figure}


	 A superconducting solution with finite gap components is obtained 
for all the doping levels considered. The gaps have opposite signs, and no 
nodes. Figs. 4(a) and (b) show these full gaps on the Fermi surfaces for 
doping $x$ = 0.082, which is qualitatively representative of all cases. 
$\Delta_{\tilde {\bf k}}^+$ is effectively constant, while $\Delta_{\tilde 
{\bf k}}^-$ has only appreciable zeroth and first harmonics, justifying the 
neglect of higher components. That the order parameters have opposite 
signs on each band corresponds within the ladder to the condition \cite{rttr} 
for ``$d$-like'' symmetry that $\Delta (k_{\perp} = 0)$ and $\Delta (k_{\perp} 
= \pi)$ have opposite signs. The order parameters are isotropic transverse 
to the ladder direction. The decrease of gap components with doping is 
quantified by $T_c$ (Fig. 5). The slower increase with decreasing doping 
below $x$ = 0.082 may be accounted for as above. Quantitatively, the values 
of $T_c$ predicted from the two-band model with spin fluctuation-mediated 
pairing approach those of the 2d planar cuprate La$_{2-x}$Sr$_x$CuO$_4$. 

\begin{figure}[hp]
\centerline{\psfig{figure=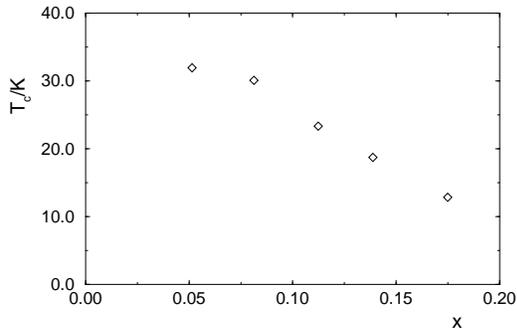,height=4.5cm,angle=270}}
\medskip
\caption{$T_c$ as a function of doping for two-band model. }
\end{figure}

	Our primary result is then that, within a specific model 
for AF spin fluctuations, the coupled ladder compound 
La$_{1-x}$Sr$_x$CuO$_{2.5}$ is predicted to be 
superconducting. Since the solutions of the gap equations have a 
full gap over the whole Fermi surface with opposite signs on the 
different sheets, there is clearly no inherent frustration for a 
superconductor with approximate $d$-wave symmetry. From this point 
of view, it appears rather that the Fermi surface form would favor 
a superconducting state. The primary weakness of the preceeding 
calculation is that it is carried out for a clean system, whereas 
in reality there are two sources of disorder. One is crystalline 
imperfections, which presumably may be controlled, and the other 
is intrinsic to the doping process which substitutes divalent 
Sr$^{2+}$ ions for trivalent La$^{3+}$ in close proximity to the 
ladders. Evidence for the disruptive influence of this random 
potential may be found in recent experiments on thin films of 
La$_{1.9}$Sr$_{0.1}$CuO$_{4}$ \cite{rlpfmsv}, in which the distance 
of Sr$^{2+}$ from the cuprate planes was varied. For greater 
separation, the transition temperature was substantially enhanced, 
from 29K to $T_c$ = 50K, and simultaneously the in-plane resistivity 
in the normal state was reduced. This source of disorder cannot be 
eliminated, and may lead to localization, particularly at lower $x$, 
which would produce a doping dependence of $T_c$ of the type manifest 
in the planar cuprates ({\it cf.} Fig. 5). The synthesis of 
crystalline samples with $x$ = 0.15 and low residual resistivities 
\cite{rmnt} is encouraging in the following regard. It provides 
hope that if high-quality single crystals could be synthesized for 
smaller $x$, there is the possibility of superconductivity, or at 
minimum of the observation of spin-gap features similar to the 
underdoped cuprates. If in the latter case a crossover to an insulating 
phase appeared at low temperatures, this would be {\it prime facie} 
evidence that indeed the presence of disorder was responsible for 
a suppression of superconductivity. 

	A related question concerns the Landau Fermi liquid nature 
of the metallic state observed at $x$ = 0.15. At first sight, the 
quasi-1d nature of La$_{1-x}$Sr$_x$CuO$_{2.5}$ should accelerate
the breakdown of Landau theory, yet it seems that the critical 
doping required to stabilize it is actually lower than in the planar 
cuprates. One possible explanation for this unexpected discrepancy 
is the role of Umklapp scattering, which has been proposed \cite{rfrs} 
as the cause of the breakdown of Landau behavior in 2d cuprate systems. 
In the present case, Umklapp scattering between the two Fermi-surface 
sheets occurs only for smaller values of $x \le x_m = 0.112$. Again 
it would be most interesting to follow the evolution of the normal 
state in samples with lower doping. 


	In conclusion, we have found that from the point of view of 
spin-fluctuation theory, the La$_{1-x}$Sr$_x$CuO$_{2.5}$ system, with 
its clear ladder structure, should favor a nodeless, $d$-wave superconducting 
state. However, this may be suppressed by the strong intrinsic disorder 
introduced by substituting Sr$^{2+}$ for La$^{3+}$. The synthesis of 
crystalline samples with low doping values would shed light on many key 
issues in the understanding of cuprates. 


We are grateful to S. Miyasaka and H. Takagi for assistance, and to the 
Swiss National Fund for financial support. BN acknowledges the generosity 
of the Treubelfonds, and DFA the support of the NSERC of Canada.

\end{document}